\documentclass{aa}
\usepackage{graphicx}
\usepackage{bm}
\usepackage{times}
\graphicspath{{./fig/}{./png/}}

\newcommand{\EQ}{\begin{equation}}
\newcommand{\EE}{\end{equation}}
\newcommand{\EQA}{\begin{eqnarray}}
\newcommand{\EEA}{\end{eqnarray}}
\newcommand{\brac}[1]{\langle #1 \rangle}
\newcommand{\pd}{\partial}

\newcommand{\CURL}{\vec{\nabla} \times }

\newcommand{\mean}[1]{\overline{#1}}
\newcommand{\meanv}[1]{\overline{\bm #1}}
\newcommand{\cst}{c_{\rm s}^2}

\newcommand{\etat}{\eta_{\rm t}}

\newcommand{\urms}{u_{\rm rms}}

\newcommand{\brms}{B_{\rm rms}}
\newcommand{\Beq}{B_{\rm eq}}
\newcommand{\Ma}{\rm Ma}

\newcommand{\kef}{k_{\rm f}}
\newcommand{\tauc}{\tau_{\rm c}}

\newcommand{\St}{{\rm St}}
\newcommand{\Sh}{{\rm Sh}}
\newcommand{\Pm}{{\rm Pm}}
\newcommand{\Rm}{{\rm Rm}}
\newcommand{\Pra}{{\rm Pr}}
\newcommand{\Ra}{{\rm Ra}}

\newcommand{\Rey}{{\rm Re}}
\newcommand{\Co}{{\rm Co}}

{}
\def\onethird{{\textstyle{1\over3}}}
\def\onehalf{{\textstyle{1\over2}}}

\begin{document}

\authorrunning{K\"apyl\"a et al.}
\titlerunning{Open and closed boundaries in convective dynamos}

   \title{Open and closed boundaries in large-scale convective dynamos}

   \author{P. J. K\"apyl\"a
          \inst{1,2}
          \and
          M. J. Korpi
          \inst{1,2}
          \and
          A. Brandenburg
          \inst{2,3}
          }

   \offprints{\email{petri.kapyla@helsinki.fi}
              }

   \institute{Department of Physics, Gustaf H\"allstr\"omin katu 2a 
              (PO Box 64), FI-00014 University of Helsinki, Finland
         \and NORDITA, Roslagstullsbacken
              23, SE-10691 Stockholm, Sweden
         \and Department of Astronomy, Stockholm University, SE-10691
              Stockholm, Sweden}

   \date{Received 23 November 2009 / Accepted 26 May 2010}

   \abstract{Earlier work has suggested that large-scale dynamos can 
   reach and maintain
   equipartition field strengths on a dynamical time scale only
   if magnetic helicity of the fluctuating
   field can be shed from the domain through open boundaries.}%
   {Our aim is to test this scenario in convection-driven dynamos
     by comparing results for open and closed boundary conditions.}
   {Three-dimensional numerical simulations of turbulent compressible
     convection with shear and rotation are used to study the effects
     of boundary conditions on the excitation and saturation of
     large-scale dynamos. Open (vertical-field) and closed 
     (perfect-conductor) boundary conditions are used for the magnetic
     field. The shear flow is such that the contours of shear are
     vertical, crossing the outer surface,
     and are thus ideally suited for driving a shear-induced magnetic
     helicity flux.}%
   {We find that for given shear and rotation rate, the growth rate of the
     magnetic field is larger if open boundary conditions are
     used. The growth rate first increases for small magnetic Reynolds
     number, $\Rm$, but then levels off at an approximately constant
     value for intermediate values of $\Rm$. For large enough $\Rm$,
     a small-scale dynamo is excited and the growth rate of the
     field in this regime increases as $\Rm^{1/2}$.
     Regarding the nonlinear regime, the saturation level of the energy of
     the total magnetic field is independent of $\Rm$ when open boundaries are
     used. In the case of perfect conductor boundaries, the saturation
     level first increases as a function of $\Rm$, but then
     decreases proportional to $\Rm^{-1}$ for $\Rm\ga 30$, indicative of
     catastrophic quenching. These
     results suggest that the shear-induced magnetic helicity flux is
     efficient in alleviating catastrophic quenching when open
     boundaries are used. The horizontally averaged mean field
     is still weakly decreasing as a function of $\Rm$ even for open
     boundaries.}%
   {}%


   \maketitle

\section{Introduction}
The classical view of the effect of shear in large-scale dynamos is
that it promotes the generation of magnetic fields, and that it is instrumental
in exciting oscillatory solutions (e.g.\ Parker \cite{P55}; Steenbeck
\& Krause \cite{SK69}). While these aspects remain unchanged, a much
more subtle effect of shear has been discovered in studies of the
saturation mechanism of the large-scale dynamo. If the
saturation of the dynamo occurs due to magnetic helicity
conservation, a fully periodic or closed system experiences
catastrophic quenching with the energy of the large-scale magnetic 
field decreasing in inverse proportion to the magnetic Reynolds 
number (Vainshtein \& Cattaneo \cite{VC92}; Cattaneo \& Hughes \cite{CH96};
Brandenburg \cite{B01}). A way to avoid
catastrophic quenching is to allow magnetic helicity fluxes to escape
from the system through open boundaries, thus allowing the large-scale
fields to saturate at equipartition strength in a dynamical time scale
(Blackman \& Field \cite{BF00}; Kleeorin et al.\ \cite{KMRS00}).
One of the most promising mechanisms, introduced by Vishniac \& Cho
(\cite{VC01}), operates by driving a flux of magnetic helicity
along the isocontours of shear.
Forced turbulence simulations with shear have confirmed that the
quenching of the $\alpha$-effect is up to 30 times weaker when closed
boundaries are replaced by open ones (Brandenburg \& Sandin \cite{BS04}).

Recent numerical experiments with forced turbulence
(Brandenburg \cite{B05}; Yousef et al.\
\cite{YHSea08a,YHSea08b}; Brandenburg et al.\ \cite{BRRK08}) and
convection (K\"apyl\"a et al.\ \cite{KKB08}, hereafter Paper~I; Hughes
\& Proctor \cite{HP09}; see also K\"apyl\"a et al.\ \cite{KKB10}) with 
imposed large-scale shear flows have
established the existence of a robust large-scale dynamo. The origin
of this dynamo is still under debate (shear--current versus incoherent
$\alpha$--shear effects).
This is discussed in detail elsewhere 
(e.g.\ Brandenburg et al.\ \cite{BRRK08}; K\"apyl\"a et al.\ \cite{KKB09b}). 
In forced turbulence simulations with fully periodic boundaries,
that do not allow magnetic helicity fluxes, large-scale magnetic
fields show slow saturation on a diffusive timescale (Brandenburg \cite{B01}).
Earlier results from convection also suggest that no appreciable
large-scale magnetic fields are seen if magnetic he\-li\-ci\-ty fluxes are
suppressed (Paper~I).

In most of the simulations of Paper~I the isocontours of shear were vertical
and thus intersecting the boundaries.
However, the boundary conditions (bc) imposed on the magnetic 
field on the vertical borders
would either allow (open vertical-field bc) or
inhibit (closed perfect-conductor bc) a net flux
out of the domain. It was shown that when
the flux is suppressed, only weak large-scale magnetic fields occur
in the saturated state at large $\Rm$.
Conversely, allowing a flux produced a significant large-scale
field. Furthermore, using a similar setup as in Tobias et al.\
(\cite{TCB08}), where the isocontours of shear are
horizontal, a similar effect was found. Thus it would appear that the
shear-induced flux plays a critical role in exciting a large-scale
dynamo in turbulent convection.
This has given rise to the impression that the dynamos seen
in Paper~I are solely due to the magnetic helicity flux and thus inherently
nonlinear (Hughes \& Proctor \cite{HP09}). In the present paper we
argue that this interpretation is incorrect and that the lack of 
appreciable large-scale magnetic fields 
in some of our earlier cases with vanishing
helicity flux is due to the fact that the critical dynamo number for the
excitation of the dynamo is larger in that case.
We also note that, more recently, large-scale dynamos have been found in
turbulent convection without shear (K\"apyl\"a et al.\ \cite{KKB09a}),
irrespective of the presence of magnetic helicity fluxes.

The origin of the dynamos reported in Paper~I was interpreted in the
context of turbulent dynamo theory in K\"apyl\"a et al.\
(\cite{KKB09b}). An important feature of turbulent dynamos is that
according to standard mean-field theory neither the turbulent
transport coefficients nor the growth rate of the field should depend
on the microscopic resistivity. The former prediction was confirmed by
K\"apyl\"a et al.\ (\cite{KKB09b}), but the latter has not yet been
studied in detail in direct simulations. This is one of the goals of
the present paper.

Another aim is to study the saturation
behaviour of the large-scale dynamo with both open and closed boundary
conditions. Firstly, in homogeneous systems, i.e.\ where the turbulent
transport coefficients do not vary in space, 
full saturation is expected to
occur on a slow diffusive time scale with closed boundaries 
whereas with open boundaries the
saturation is expected to happen on a dynamical time scale if a 
helicity flux is present (Brandenburg \cite{B01}).
However, in inhomogeneous systems the situation is likely to be 
more complex (e.g.\ Mitra et al.\ \cite{MCCTB10}).
Furthermore, simulations of helically forced isotropic
turbulence have shown that, in the absence of shear, the saturation
level of the magnetic field scales as
$\meanv{B}^2/\Beq^2\propto\Rm^{-1}$ for open boundaries
(Fig.~2 of Brandenburg \& Subramanian \cite{BS05a}).
When shear is present, the shear-induced magnetic
helicity flux should alleviate the $\Rm$-dependence of the saturation
level. Preliminary results in Paper~I indicate that this might indeed be
true, but the results are not fully conclusive because other parameters, such
as the Rayleigh number and the fluid Reynolds number changed when
$\Rm$ was varied. In the present paper we perform a more detailed
study where only $\Rm$ is varied.

\section{Model and methods}
\subsection{Setup and boundary conditions}
We use the same setup as in Paper~I in which a small rectangular portion of a
star is modelled by a box situated at colatitude $\theta$. The
coordinate system is such that $(x,y,z)$ correspond to
$(\theta,\phi,r)$ in a spherical coordinate system.  The dimensions of
the domain are $(L_x, L_y, L_z) = (4,4,2)d$, where $d$ is the depth of
the convectively unstable layer, which is also used as our unit
length. The box is divided into three layers, an upper cooling layer,
a convectively unstable layer, and a lower stable overshoot layer.
Gravity points in the negative $z$ direction, $\bm{g}=(0,0,-g)$,
rotation is in the positive $z$ direction, $\bm{\Omega}=(0,0,\Omega)$, i.e.\
$\theta=0$ corresponding to the north pole,
and shear is in the $y$ direction, with $\meanv{U}=(0,Sx,0)$, where
$S$ is the shear rate.

The values $(z_1, z_2, z_3, z_4) = (-0.85, 0, 1, 1.15)d$ give the
vertical positions of the bottom of the box, the bottom and top of the
convectively unstable layer, and the top of the box, respectively. We
use a heat conductivity profile such that the associated hydrostatic
reference solution is piecewise polytropic with indices $(m_1, m_2,
m_3)=(3, 1, 1)$. We apply a cooling term in the thin uppermost layer
which makes this region
nearly isothermal and hence stably stratified.  The bottom layer is
also stably stratified, and the middle layer is convectively unstable.

At the vertical boundaries we use stress-free boundary conditions for 
the velocity,
\begin{equation}
U_{x,z} = U_{y,z} = U_z = 0.
\end{equation}
For the magnetic field either vertical-field or perfect-conductor
boundary conditions are used.
Occasionally we refer to them also as open and closed, respectively.
Thus, we have
\begin{eqnarray}
B_x = B_y &=& 0 \quad\mbox{(vertical-field, VF, or open)},\\
B_{x,z} = B_{y,z} = B_z &=& 0 \quad\mbox{(perfect-conductor, PC, closed)},
\label{magbcs}
\end{eqnarray}
the former allowing magnetic helicity flux while the latter one does not.
This also motivates the usage of the names `open' and `closed' for the
two boundary conditions.
The $y$-direction is periodic and shearing-periodic conditions are
used in the $x$-direction (e.g.\ Wisdom \& Tremaine \cite{WT88}). 
A constant temperature gradient is maintained
at the base which leads to a steady influx of heat due to the constant
heat conductivity.

The simulations were performed with the {\sc Pencil Code}%
\footnote{\texttt{http://pencil-code.googlecode.com/}},
which uses sixth-order explicit finite differences in space and 
third-order accurate time stepping method.
Resolutions of up to $512^3$ mesh points were used.

\begin{table*}[t]
\centering
\caption[]{Summary of the runs with varying $\Pm$.
  }
      \label{tab:runs}
      \vspace{-0.5cm}
     $$
         \begin{array}{p{0.05\linewidth}ccrccccccccc}
           \hline
           \noalign{\smallskip}
Run & $grid$ & \Ma   & \Pm & \Rm & \tilde\lambda & \epsilon_{\rm f} & k_{\rm f}^{(\omega)}/\kef & \tilde\brms & \tilde{\mean{B}}_x & \tilde{\mean{B}}_y & \tilde{\mean{B}} & $bc$ \\ \hline 
VF1 & 128^3  & 0.049 & 0.05 & 0.78& \!\!\!-0.027 & -0.070 & 1.63 &   -           &   -                &     -              &   -     &  $VF$ \\ 
VF2 & 128^3  & 0.034 & 0.10 & 1.1 &  0.013 & -0.073 & 1.58 & 1.04        & 0.14               &    0.99            & 1.00    &  $VF$ \\ 
VF3 & 128^3  & 0.041 & 0.17 & 2.2 &  0.030 & -0.067 & 1.58 & 2.04        & 0.19               &    1.98            & 1.99    &  $VF$ \\ 
{\bf VF4} & 128^3  & 0.046 & 0.25 & {\bf 3.7} & 0.036 & -0.066 & 1.60 & 2.28        & 0.22               &    2.18            & 2.18    &  $VF$ \\ 
VF5 & 128^3  & 0.046 & 0.50 & 7.4 & 0.042 & -0.066 & 1.60 & 2.02        & 0.19               &    1.85            & 1.85    &  $VF$ \\ 
VF6 & 128^3  & 0.044 & 1.00 & 14  & 0.046 & -0.068 & 1.63 &  1.94        & 0.22               &    1.65            & 1.66    &  $VF$ \\ 
{\bf VF7}& 128^3  & 0.042 & 2.50 & {\bf 34} & 0.038 & -0.068 & 1.60 & 2.06        & 0.22               &    1.62            & 1.63    &  $VF$ \\ 
VF8 & 256^3  & 0.040 & 5.00 & 63  &  0.032 & -0.069 & 1.63 & 2.13        & 0.20               &    1.57            & 1.58    &  $VF$ \\ 
VF9 & 256^3  & 0.039 & 10.00 & 122 &  0.047 & -0.062 & 1.64 & 2.09        & 0.20               &    1.31            & 1.32    &  $VF$ \\ 
{\bf VF10}& 512^3  & 0.035 & 20.00 & {\bf 222}& 0.071 & -0.067 & 1.68 & 2.14        & 0.21               &    1.09            & 1.11    &  $VF$ \\ 
           \hline
PC1 & 128^3  & 0.045 & 0.25 & 3.6 & \!\!\!-0.019 & -0.065 & 1.61 &  -         &  -                 &     -              &  -      &  $PC$ \\ 
{\bf PC2}& 128^3  & 0.034 & 0.50 & {\bf 5.4}& 0.011 & -0.072 & 1.57 & 1.08        & 0.14               &    0.92            & 0.93    &  $PC$ \\ 
PC3 & 128^3  & 0.035 & 0.67 & 7.5 & 0.016 & -0.072 & 1.58 & 1.67        & 0.20               &    1.49            & 1.50    &  $PC$ \\ 
PC4 & 128^3  & 0.040 & 1.00 & 13  & 0.021 & -0.076 & 1.58 & 2.44        & 0.26               &    2.23            & 2.23    &  $PC$ \\ 
{\bf PC5}& 128^3  & 0.040 & 2.50 & {\bf 32} & 0.028 & -0.072 & 1.58 & 3.22        & 0.31               &    2.90            & 2.90    &  $PC$ \\ 
PC6 & 256^3  & 0.038 & 5.00 & 61  & 0.027 & -0.062 & 1.62 & 3.05        & 0.33               &    2.55            & 2.56    &  $PC$ \\ 
PC7 & 256^3  & 0.035 & 10.00& 112 & 0.049 & -0.065 & 1.63 & 2.29        & 0.22               &    1.53            & 1.55    &  $PC$ \\ 
{\bf PC8}& 384^3  & 0.037 & 14.29& {\bf 168}& 0.057 & -0.056 & 1.65 & 1.83        & 0.17               &    0.95            & 0.96    &  $PC$ \\ 
PC9 & 512^3  & 0.038 & 20.00 & 239 & 0.064  &-0.054 & 1.64 &     1.62   &           0.16     &          0.74    &  0.76   &  $PC$ \\ 
           \hline
         \end{array}
     $$
\tablefoot{
$\tilde\lambda$, $\epsilon_{\rm f}$, and $\kef^{(\omega)}$ are given 
for the kinematic stage whereas all the other numbers are given for 
the saturated state of the dynamo. The last column specifies the 
vertical boundary condition for the magnetic field. In all runs, 
$\Co\approx0.8$, $\Sh\approx-0.4$, $\Rey\approx12$, and 
$\Pra\approx0.7$. The adiabatic sound speed squared varies in the 
range $\cst/(gd)=0.33-1.38$ in the domain. Six of the runs are 
discussed in particular detail, so we have marked them here in 
boldface and have also highlighted the corresponding values of $\Rm$.
}
\end{table*}

\subsection{Units, nondimensional quantities, and parameters}

Dimensionless quantities are obtained by setting
\begin{eqnarray}
d = g = \rho_0 = c_{\rm P} = \mu_0 = 1\;,
\end{eqnarray}
where $g$ is the constant acceleration due to gravity, $\rho_0$ is the
density at $z_2$, $c_{\rm P}$ is the heat capacity at constant
pressure, and $\mu_0$ is the vacuum permeability. The units of length,
time,
velocity, density, specific entropy, and magnetic field are then
\begin{eqnarray}
&& [x] = d\;,\;\; [t] = \sqrt{d/g}\;,\;\; [U]=\sqrt{dg}\;,\;\; [\rho]=\rho_0\;,\;\; \nonumber \\ && [s]=c_{\rm P}\;,\;\; [B]=\sqrt{dg\rho_0\mu_0}\;. 
\end{eqnarray}
The system is controlled by the Prandtl, Reynolds, and Rayleigh numbers,
\begin{eqnarray}
\Pra&=&\frac{\nu}{\chi_0}\;,\;\; \Pm=\frac{\nu}{\eta}\;,\\
\Rey&=&\frac{\urms}{\nu \kef}\;,\;\;
\Rm=\frac{\urms}{\eta \kef}\equiv\Pm\,\Rey\;, \;\; \\ 
\Ra&=&\frac{gd^4}{\nu \chi_0} \bigg(-\frac{1}{c_{\rm P}}
\frac{{\rm d}s}{{\rm d}z} \bigg)_{z_{\rm m}}\;,
\end{eqnarray}
where $\nu$ is the kinematic viscosity, $\chi_0 = K/(\rho_{\rm m}
c_{\rm P})$ is the thermal diffusivity with $\rho_{\rm m}$ the
density at the middle of the convectively unstable layer at $z_{\rm
  m}=\onehalf(z_3-z_2)$,
$\eta$ is the magnetic diffusivity, $\urms$ the root mean
square velocity, and $\kef=2\pi/d$ is an estimate of the wavenumber of
the energy-carrying eddies. The entropy gradient, measured at $z_{\rm
  m}$ in the non-convecting initial state, is given by
\begin{eqnarray}
-\frac{1}{c_{\rm P}}\frac{{\rm d}s}{{\rm d}z}\bigg|_{z_{\rm m}} = \frac{\nabla-\nabla_{\rm ad}}{H_{\rm P}}\;,
\end{eqnarray}
with $\nabla_{\rm ad} = 1-1/\gamma$ and $\nabla = (\pd \ln T/\pd \ln
p)_{z_{\rm m}}$, where $\gamma=c_{\rm P}/c_{\rm V}=5/3$ is the ratio
of the heat capacities at constant pressure and volume, respectively, and
$H_{\rm P}$ is the pressure scale height, also at $z=z_{\rm m}$.
The differentials of logarithmic temperature and entropy are related via
${\rm d}s/c_{\rm P}={\rm d}\ln T-\nabla_{\rm ad}{\rm d}\ln p$, and
${\rm d}\ln p={\rm d}\ln T+{\rm d}\ln\rho$.

The degree of stratification is controlled by the parameter
\begin{eqnarray}
\xi_0 = \frac{(\gamma-1) e_0}{gd}\;,
\end{eqnarray}
where $e_0$ is the internal energy at $z=z_4$. 
The internal energy is related to the temperature via $e=c_{\rm V}T$.
We use $\xi_0=1/3$ in all models, which yields a density contrast of
about 27 over the whole domain.

The strengths of rotation and shear are quantified by the Coriolis and
shear numbers
\begin{eqnarray}
\Co = \frac{2\Omega}{\urms \kef}\;,\;\; \Sh = \frac{S}{\urms \kef},
\end{eqnarray}
where $(\urms \kef)^{-1}$ gives an estimate of the convective turnover
time.

\begin{table}[t]
\centering
\caption[]{Summary of the runs with varying $\Co$ and $\Sh$.}
      \label{tab:runs2}
      \vspace{-0.5cm}
     $$
         \begin{array}{p{0.075\linewidth}cccccc}
           \hline
           \noalign{\smallskip}
Run & $grid$ & \Co   & \tilde\lambda & \epsilon_{\rm f} & k_{\rm f}^{(\omega)}/\kef & $bc$ \\ \hline
VF11 & 128^3  & 0.07 & 0.000 & -0.014 & 1.29 & $VF$ \\ 
VF12 & 128^3  & 0.16 & 0.006 & -0.022 & 1.40 & $VF$ \\ 
VF13 & 128^3  & 0.34 & 0.023 & -0.041 & 1.52 & $VF$ \\ 
VF14 & 128^3  & 0.63 & 0.038 & -0.068 & 1.60 & $VF$ \\ 
VF15 & 128^3  & 1.36 & 0.054 & -0.097 & 1.76 & $VF$ \\ 
           \hline
PC9  & 128^3  & 0.07 &\!\!\!-0.008 & -0.012 & 1.30 & $PC$ \\ 
PC10 & 128^3  & 0.16 & 0.003 & -0.022 & 1.41 & $PC$ \\ 
PC11 & 128^3  & 0.38 & 0.014 & -0.042 & 1.50 & $PC$ \\ 
PC12 & 128^3  & 0.75 & 0.028 & -0.072 & 1.58 & $PC$ \\ 
PC13 & 128^3  & 1.55 & 0.034 & -0.098 & 1.74 & $PC$ \\ 
           \hline
         \end{array}
     $$ 
\tablefoot{
Here ${\rm Ma}\approx0.045$, $\Rm\approx35$, $\Pm=2.5$, and 
$\Sh=-\onehalf\Co$ (i.e.\ $S=-\Omega$) in all runs. All numbers 
are given for the kinematic state.
}
\end{table}

\subsection{Characterization of the simulation data}

Our runs can be characterized using the following quantities.
We express the rms velocity in the unit of $(gd)^{1/2}$,
which is related to the Mach number, and thus define
\begin{equation}
\mbox{Ma}=\urms/(gd)^{1/2}.
\end{equation}
The normalized growth rate of magnetic field is expressed as
\begin{equation}
\tilde\lambda=\lambda/(\urms\kef).
\end{equation}
The typical wavenumber of the energy-carrying eddies and
the relative kinetic helicity are given by
\begin{equation}
\kef^{(\omega)}=\omega_{\rm rms}/\urms,
\quad
\epsilon_{\rm f}=\brac{\bm{\omega}\cdot\bm{u}}/(\brac{\urms} 
\brac{\omega_{\rm rms}}),
\end{equation}
where $\bm\omega=\CURL\bm{u}$ is the vorticity and
the angular brackets denote volume averaging.
The normalized rms magnetic field strength is given by
\begin{equation}
\tilde\brms=\brms/\Beq,
\end{equation}
where $\Beq=\sqrt{\mu_0 \brac{\rho \bm{U}^2}}$ is the equipartition 
value of the field.
Corresponding values for the horizontally averaged (denoted by an overbar) 
horizontal components of the field are given by
\begin{equation}
\tilde{\mean{B}}_x=\langle \mean{B}_x^2\rangle^{1/2}/\Beq,\quad
\tilde{\mean{B}}_y=\langle \mean{B}_y^2\rangle^{1/2}/\Beq,
\end{equation}
and $\tilde{\mean{B}}=(\tilde{\mean{B}}_x^2+\tilde{\mean{B}}_y^2)^{1/2}$,
is referred to as the total mean field.

Error bars are estimated by dividing the time series into three equally
long parts and computing a mean value individually for each of
these. The largest deviation from the average over the full
time series is taken to represent the error.

\section{Results}
In order to study the effects of different boundary conditions and
magnetic helicity fluxes on the excitation and saturation of
convective dynamos, we perform a number of simulations with 
perfect-conductor and vertical-field boundary conditions. 
We follow two lines in parameter space where we either
vary $\Pm$ and $\Rm$, keeping all other parameters fixed, or we fix
both Reynolds numbers and vary $\Co$ and $\Sh$. The runs
are listed in Tables~\ref{tab:runs} and \ref{tab:runs2}.

\subsection{Dynamo excitation}
We begin by investigating the linear regime where the magnetic 
field is still weak.
In that case the values of $u_{\rm rms}$ are not yet quenched by the
field, and so the resulting values of $\Rm$ are somewhat larger than in
the saturated case.
Turbulent dynamo theory predicts that large-scale dynamo action is
possible if the magnetic Reynolds number and a relevant dynamo number
exceed critical values that depend on the details of the system in
question (e.g.\ Moffatt \cite{M78}; Krause \& R\"adler \cite{KR80};
R\"udiger \& Hollerbach \cite{RH04}). Furthermore,
boundary conditions also
influence the excitation of large-scale dynamos
(e.g.\ Choudhuri \cite{Cho84}; Jouve et
al.\ \cite{Jea08}).
In order to study these effects,
we determine the critical $\Rm$ and the critical dynamo number, here
implicitly described by the strengths of rotation and shear, needed to
excite the dynamo with both open and closed boundary conditions.

\subsubsection{Dependence on $\Rm$} 
\begin{figure}[t]
\centering
\includegraphics[width=\columnwidth]{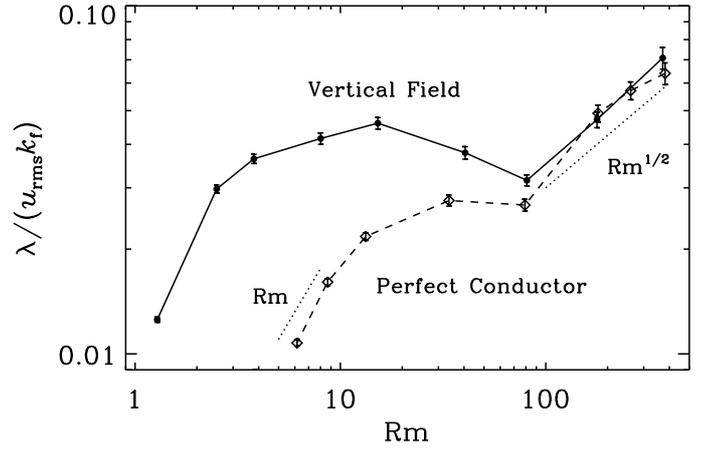}
\caption{Growth rate $\lambda$ of the magnetic field as a
  function of $\Rm$ from the runs listed in
  Table~\ref{tab:runs}. Solid (dashed) line with closed (open) symbols
  shows the results for the VF (PC) boundary conditions. 
  Dotted lines proportional to
  $\Rm$ and $\Rm^{1/2}$ are shown for reference.}
\label{fig:pgr}
\end{figure}

From the perspective of mean-field dynamo theory a,
large-scale dynamo should operate if $\Rm\ga1$.
We start the simulations with a random small-scale magnetic field of
the order of $10^{-4}\Beq$ and monitor the time evolution of the total
magnetic field quantified by $\brms$.
We indeed find that in runs with vertical-field boundary conditions 
the dynamo is excited
for $\Rm\ga1$, whereas in the case of perfectly conducting
boundaries, the critical value is about five times higher (see
Table~\ref{tab:runs} and Fig.~\ref{fig:pgr}).

As a consistency check we use a one-dimensional mean-field model
written in terms of the vector potential
\begin{eqnarray}
\dot{\mean{A}}_i=-\mean{U}_{j,i}\mean{A}_j+\alpha_{ij}\mean{B}_j
-(\eta_{ij}+\eta\delta_{ij})\mu_0\mean{J}_j
\end{eqnarray}
where $\dot{\mean{A}}_i$ denotes a time derivative and
$\mean{U}_{j,i}\mean{A}_j=(S\mean{A}_y,0,0)$ is the shear term.  The
mean magnetic field is given by
$\mean{\bm{B}}=(-\mean{A}_y',\mean{A}_x',0)$, the mean current density
is given by $\mu_0\mean{J}_i=-\mean{A}_i''$, and primes denote
$z$-derivatives. The coefficients $\alpha_{ij}$ and $\eta_{ij}$,
describing the turbulent transport coefficients relevant for
$z$-dependent large-scale fields, are taken from a test field
calculation (Run~D4 of K\"apyl\"a et al.\ \cite{KKB09b}; see their
Fig.\ 13) which is the same as our run VF7, but with $\Pm=5$ instead of
$2.5$ used here. We then use either VF or PC boundary conditions,
defined by Eq.~(\ref{magbcs}), and determine the marginal value of
$\eta$ that allows dynamo action. This is quantified by a Reynolds
number $\Rm_{\rm c}=\urms/(\eta_{\rm c} \kef)$, where $\urms$ is taken
from the same run as the turbulent transport coefficients and
$\eta_{\rm c}$ is the marginal value of the microscopic magnetic
diffusivity.

We find a marginal value $\Rm_{\rm c}\approx 1.0$ for the vertical-field
boundary conditions, whereas in the case of perfect-conductor boundaries
$\Rm_{\rm c}\approx4.1$ is obtained. These values are very close to
those obtained from direct simulations, although one must bear in
mind that the $\Rm$-dependence of $\alpha_{ij}$ and $\eta_{ij}$ at
small values of $\Rm$ (K\"apyl\"a et al.\ \cite{KKB09b}) is not taken
into account in our mean-field modelling. However, the qualitative
result that the dynamo is harder to excite when perfect-conductor
boundaries are used is consistent with the direct simulations.

We measure the growth rate, $\lambda$, of the rms magnetic field as
\begin{equation}
\lambda = \left\langle{\rm d} \ln \brms/{\rm d}t\right\rangle_{\rm t},
\end{equation}
where $\langle \ldots \rangle_{\rm t}$ denotes a time average
over the kinematic range where $\brms$ grows exponentially.
We find that, for Reynolds numbers close to the marginal value, $\lambda$
increases roughly proportional to $\Rm$ (see Fig.~\ref{fig:pgr}) in 
accordance with theoretical
considerations. However, soon this increase ceases and $\lambda$
levels off to a roughly constant value for intermediate values of
$\Rm$. For
$\Rm\ga60$ the growth rate is consistent with
$\lambda\propto\Rm^{1/2}$. This behaviour can be understood as follows:
for small $\Rm$ the large-scale dynamo is excited, but the growth rate is
still affected by the relatively large value of the microscopic 
magnetic diffusivity $\eta$. 
In the present system the small-scale dynamo is
excited when $\Rm\ga30$, but this contribution begins to dominate over
the growth rate of the large-scale field only for $\Rm\ga60$ where our
data for $\lambda$ is consistent with an $\Rm^{1/2}$ scaling. This
is in accordance with findings of 
Schekochihin et al.\ (\cite{Scheko04}) and Haugen et al.\ (\cite{{HBD04}})
for forced isotropic turbulence.

\begin{figure}[t]
\centering
\includegraphics[width=\columnwidth]{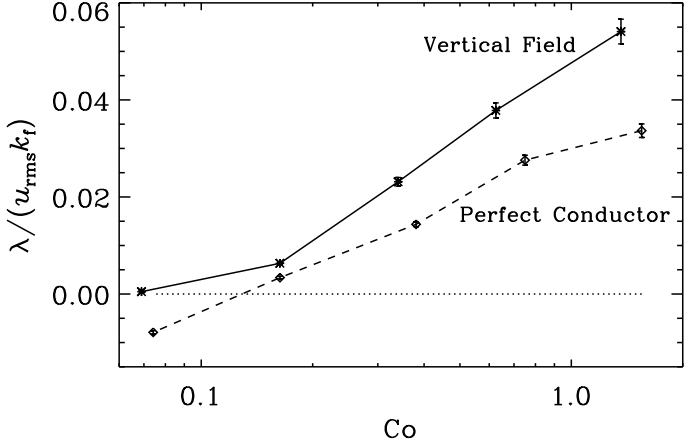}
\caption{Growth rate $\lambda$ of the magnetic field as a
  function of $\Co$ with open (VF, solid line) and closed (PC, dashed
  line) boundary conditions for the runs listed in Table~\ref{tab:runs2}.}
\label{fig:pgr_sh}
\end{figure}

\begin{figure}[t]
\centering
\includegraphics[width=\columnwidth]{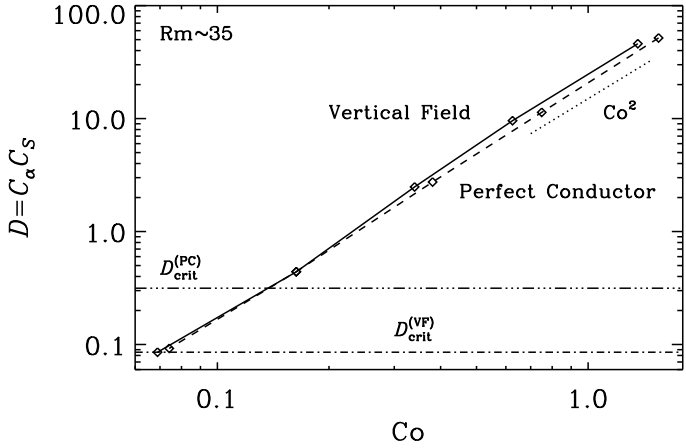}
\caption{Dynamo numbers according to Eq.~(\ref{dynn}) as functions of
  rotation for the runs listed in Table~\ref{tab:runs2}. The solid
  (dashed) curve shows the results for vertical-field 
  (perfect-conductor) boundary conditions. The horizontal dash-dotted and
  dash-triple-dotted lines indicate the critical dynamo number in the
  case of open and closed boundaries, respectively. Power law
  proportional to $\Co^2$ is shown for reference.}
\label{fig:pdynn}
\end{figure}

\subsubsection{Dependence on shear and rotation}
In addition to the dependence on the magnetic Reynolds number, 
we study the influence of shear on
the excitation of the dynamo. We use a setup where $\Rm\approx35$ in
the case $\Co=\Sh=0$, in which case the small-scale dynamo is
marginal. We then introduce shear and rotation into the system and
keep the quantity $q=-S/\Omega=1$ fixed in all cases. Our results
for the growth rate of the magnetic field in the range
$\Co\approx0.07\ldots1.5$ are shown in Fig.~\ref{fig:pgr_sh}. We find
that for a given $\Co$ the growth rate is always greater in the
simulations with open boundaries (see also Table~\ref{tab:runs2}).
The case $\Co\approx0.07$ (Run~VF11) is very
close to marginal in the case of open boundaries, but clearly
subcritical when perfectly conducting boundaries are used (Run~PC9). 
The critical value of
$\Co$ in the perfect-conductor case seems to be roughly twice as large
as the corresponding value in the vertical-field runs.
To be more precise, we estimate the relevant dynamo numbers as 
\begin{eqnarray}
C_\alpha&\equiv&\frac{\alpha}{\eta_{\rm t}k_1}\approx -\epsilon_{\rm f}\frac{\kef^{(\omega)}}{k_1},\\
C_S&\equiv&\frac{S}{\eta_{\rm t}k_1^2}\approx 3\,\Sh\left(\frac{\kef}{k_1}\right)^2,
\end{eqnarray}
where $k_1=\pi/L_z$ is the wavenumber of the leading decay mode for both
boundary conditions.
Here we approximate the $\alpha$ effect 
and turbulent diffusivity by
\begin{eqnarray}
\alpha \approx -\onethird \tau_{\rm c} \brac{\bm\omega\cdot\bm{u}},\quad
\eta_{\rm t} \approx \onethird \tau_{\rm c} \urms^2.
\end{eqnarray}
Furthermore, we assume the Strouhal number,
\begin{eqnarray}
\St=\tauc \urms \kef,
\end{eqnarray}
of unity
(as found by Brandenburg \& Subramanian \cite{BS05b}) so that $\tau_{\rm
  c}=(\urms \kef)^{-1}$. The fractional helicity is defined via 
$\epsilon_{\rm f}=\brac{\bm\omega\cdot\bm{u}}/(\urms \omega_{\rm rms})$, where
$\omega_{\rm rms}=\kef^{(\omega)} \urms$. 
We can now define the dynamo number as 
\begin{equation}
D\equiv C_\alpha C_S. \label{dynn}
\end{equation}
For the marginal open-field Run~VF11 we find a critical dynamo number 
$D_{\rm crit}^{\rm(VF)} \approx0.1$. The dynamo numbers for the perfect 
conductor Runs~PC9 and
PC10 are $0.7$ and $3.5$, respectively. Linear interpolation of
the magnetic field growth rates give a critical value of $D_{\rm
  crit}^{\rm(PC)}\approx0.3$ which is roughly three times greater than in
the vertical-field case (see Fig.~\ref{fig:pdynn}).
We find that the dynamo number is approximately proportional to
$\Co^2$ in the range of parameters studied here.
This dependence can be understood qualitatively as follows: the
$\alpha$ effect, and hence $C_\alpha$, increases approximately
proportional to $\Co$ in the range of parameters studied here (see
K\"apyl\"a et al.\ \cite{KKB09b}). Furthermore, as we keep the ratio
$S/\Omega$ constant, shear and $C_S$ also increase proportionally
to $\Co$.

In the preceding simplified analysis we have estimated
$\alpha$ and $\etat$ using volume averages that are not ideally suited
for the present case (see, e.g.\ K\"apyl\"a et al.\
\cite{KKB09b}). Furthermore, we have assumed here that the dynamo
is of pure $\alpha$-shear type and have neglected all other possible
induction
effects (shear-current, $\bm\Omega\times\meanv{J}$, and incoherent
$\alpha$-shear effects) that can assist in dynamo action 
(K\"apyl\"a et al.\ \cite{KKB09b}). This means that
the quantitative results should be taken with some amount of caution.

In Paper~I, the only case where open and closed boundaries were compared
was the one where only shear was present, i.e.\ $\Sh\neq0$, 
whereas $\Co=0$ (Run~C
with vertical-field boundary conditions versus Run~C' with 
perfect-conductor boundary conditions). In
Paper~I we also showed that, in the absence of rotation, the growth rate
of the magnetic field is reduced in comparison to the case where
both shear and rotation are present. Furthermore, our present results 
suggest that the
large-scale dynamo is harder to excite in the case of 
perfect-conductor boundaries. Thus, a reasonable explanation to the 
slow growth of
the large-scale field in Run~C' of Paper~I is that the system is
close to being marginally excited, and not, as suggested by Hughes
\& Proctor (\cite{HP09}), the result of some nonlinear effect 
arising from magnetic helicity flux out of the system.
A straightforward quantitative analysis in terms of a
classical $\alpha$--shear dynamo described above is, however, not useful in the
non-rotating case. This is shown by the results of K\"apyl\"a et
al.\ (\cite{KKB09b}): the main contribution to dynamo action in this
case is due to the incoherent $\alpha$--shear process which relies on
the fluctuations of $\alpha$ (e.g.\ Vishniac \& Brandenburg
\cite{VB97}; Sokolov \cite{S97}; Silant'ev \cite{S00}; Proctor
\cite{P07}).

\begin{figure*}[t]
\centering
\includegraphics[width=\textwidth]{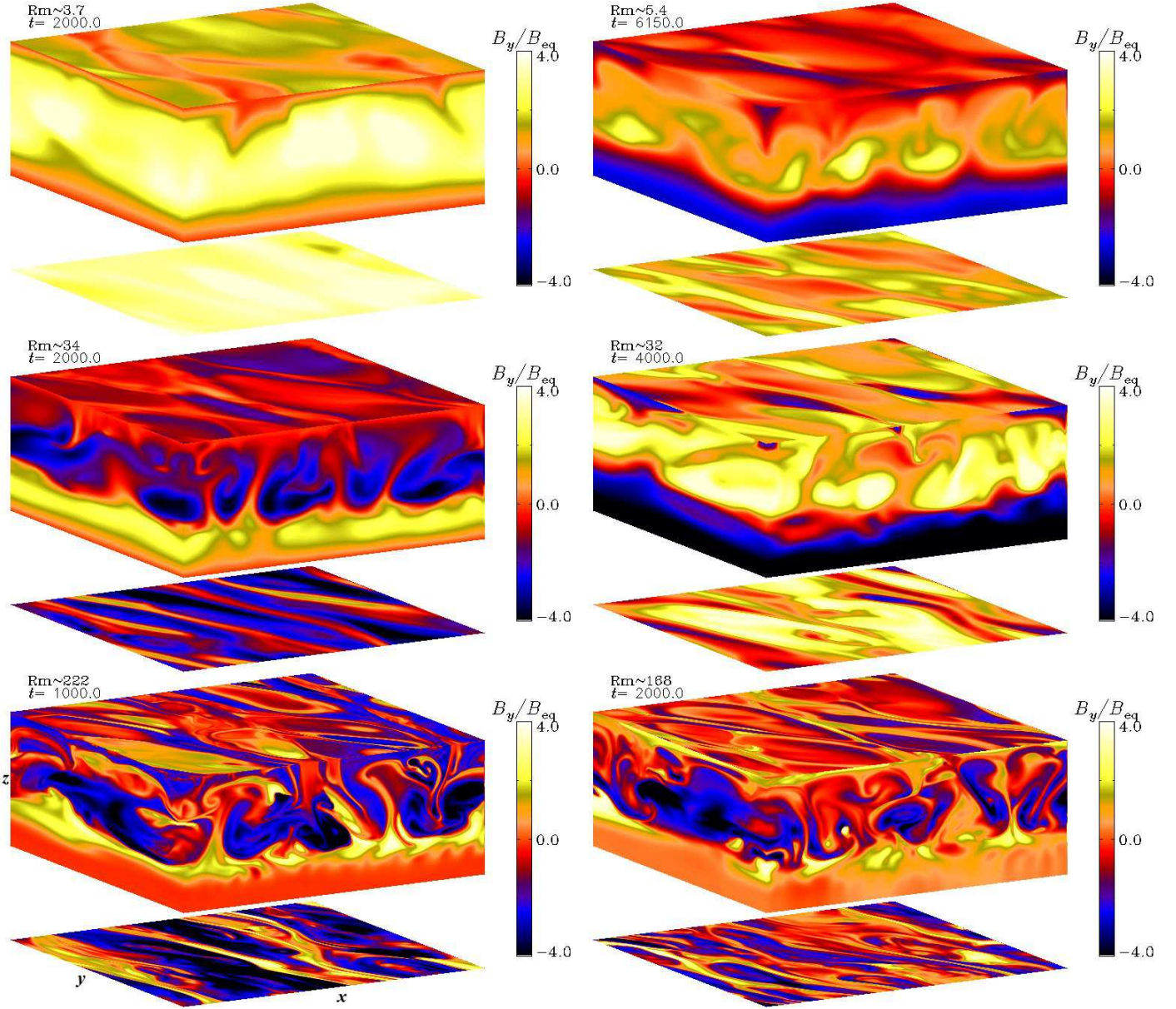}
\caption{Magnetic field component $B_y$ from runs with vertical-field
  (left column) and perfect-conductor (right column) boundary
  conditions. Three different cases with low (top), intermediate
  (middle) and high (bottom) $\Rm$ are shown. The sides of the boxes
  show the $B_y$ field at the periphery of the domain whereas the top and
  bottom panels show the field at $z=z_3$ and $z=z_2$, respectively.}
\label{fig:boxes}
\end{figure*}

\begin{figure*}[t]
\centering
\includegraphics[width=\textwidth]{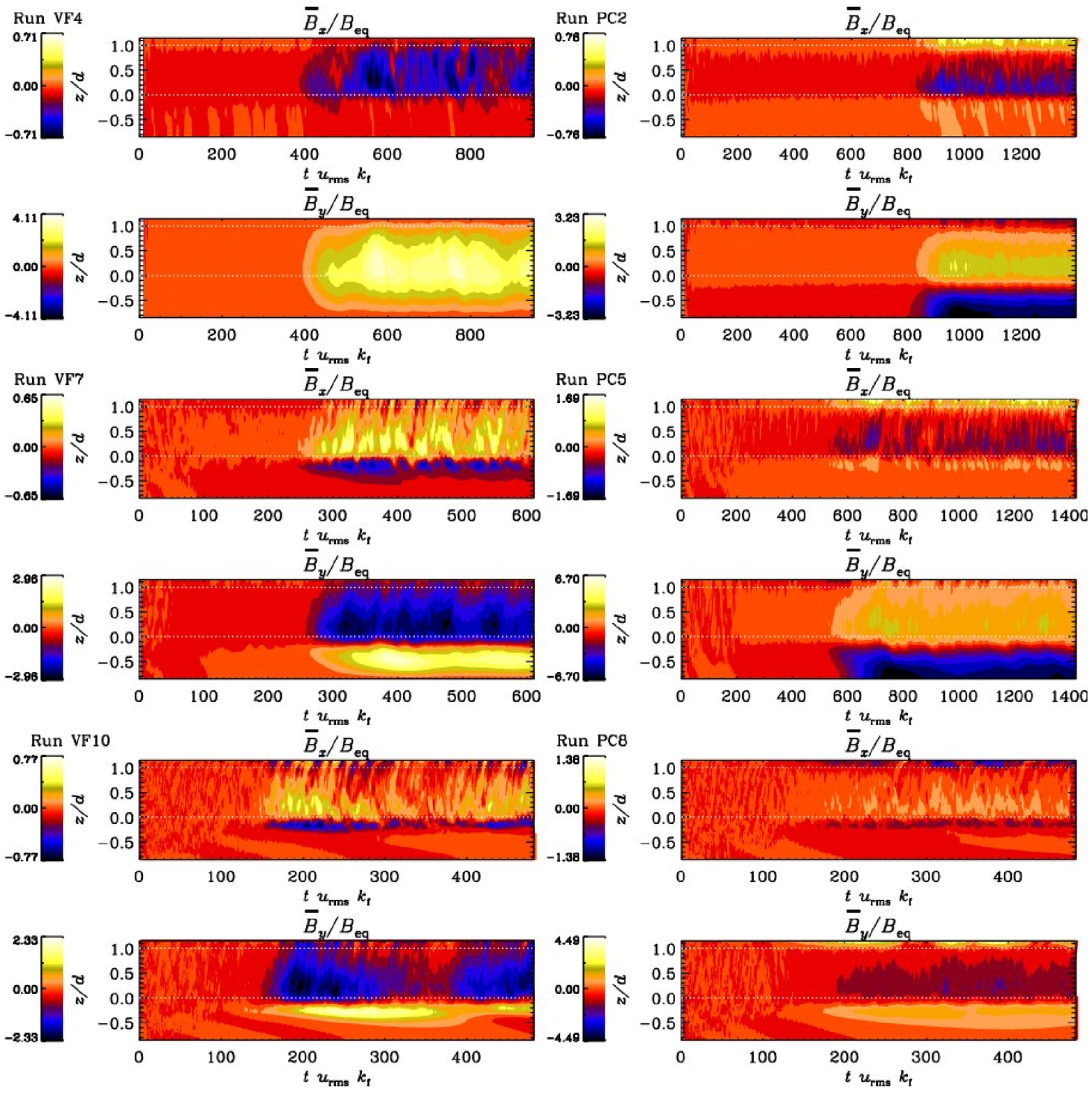}
\caption{Horizontally averaged magnetic field components $B_x$ and
  $B_y$ from the same six runs as in Fig.~\ref{fig:boxes}.
  Those runs and the corresponding values of $\Rm$ are also highlighted
  in Table~\ref{tab:runs} in bold face.
  The dotted
  white lines at $z=0$ and $z=d$ denote the bottom and top of the
  convectively unstable layer, respectively.
  The runs on the left are for open boundaries and those on the right
  are for closed ones.
  The magnetic Reynolds number increases downwards.}
\label{fig:bflys}
\end{figure*}

\subsection{Nonlinear regime}

We find that in all cases where a dynamo is excited, large-scale
magnetic fields are also generated. In the low-$\Rm$ runs the
large-scale contribution is substantial, i.e.\ $\meanv{B}/\brms$ is 
close to unity already in the kinematic
regime. In the cases where also the fluctuation dynamo is
excited, a large-scale pattern is discernible only
once the dynamo has reached saturation.
This feature is common to all known large-scale dynamos
and was already seen in simulations of Brandenburg (\cite{B01}).
Figure \ref{fig:boxes} shows visualizations of
the streamwise component of the field in the saturated state for three Reynolds
numbers, representing low, intermediate, and
high $\Rm$, with open and closed boundaries (see Table \ref{tab:runs}).
The runs depicted in Fig.~\ref{fig:boxes} are indicated in boldface
in the table.
Although an increasing amount of
small-scale features is seen with increasing $\Rm$, a large-scale
pattern is clearly visible in all cases.

Unlike the case with just rotation and no shear,
where the mean field shows variations in the horizontal directions
(K\"apyl\"a et al.\ \cite{KKB09a}), here the $k_x=k_y=0$ mode
dominates the large-scale field and thus horizontal averages are
suitable. Space-time diagrams of the horizontally averaged magnetic
field components, $\mean{B}_x$ and $\mean{B}_y$, are shown in
Fig.~\ref{fig:bflys} for the same runs as in Fig.~\ref{fig:boxes}.
We find that the large-scale field is non-oscillatory in all cases,
which is in agreement with earlier results (Paper~I; Hughes \& 
Proctor \cite{HP09}; K\"apyl\"a et al.\ \cite{KKB10}).
In most cases, regardless of the
boundary conditions, the field changes sign near the base of the
convectively unstable layer, with the exception of the runs with
vertical-field boundary conditions (VF runs) with the lowest values of $\Rm$.
(The solutions are invariant under sign reversal, $\bm{B}\to-\bm{B}$, and
both realizations of the large-scale field are found in Figs.~\ref{fig:boxes}
and \ref{fig:bflys}, depending just on the initial conditions.)
In the perfect-conductor runs (PC runs), the
field near the top of the domain has a different sign than in the bulk
of the convection zone, whereas in the VF runs such
behaviour is not seen.
We find that in the PC runs the layer of oppositely
directed field becomes progressively thinner as $\Rm$ is increased,
leading to strong gradients near the boundary (see
Fig.~\ref{fig:bflys}). This can possibly explain the numerical
problems encountered in the $\Rm\approx239$ simulation (Run~PC9) with
perfect-conductor boundary conditions.
The saturated state of Run~PC9 is significantly shorter
than that of the other runs due to numerical instability that
prevented running the simulation further.

It turns out that the emergence of the large-scale field occurs
progressively earlier as $\Rm$ is increased.
This can be seen most clearly in Fig.~\ref{fig:bflys}, which shows that
the large-scale field has reached saturation in less than 200 turnover times
for $\Rm=222$ (Run~VF10), while for Runs~VF7 and VF4 with $\Rm=34$ and
3.7, the saturation times exceed 300 and 400 turnover times, respectively.
A similar trend is seen also in the PC runs.
In an earlier study we found that the mean values of the turbulent
transport coefficients relevant for the generation of large-scale
magnetic fields, $\alpha_{yy}$ and $\etat$, remain constant within the
errors given that $\Rm\ga8$; see Fig.~5 of K\"apyl\"a et al.\
(\cite{KKB09b}). However, as the magnetic Reynolds number is
increased, the fluctuations of $\alpha$ tend to increase as well. Such
fluctuations can contribute to the incoherent $\alpha$-shear process
which can possibly explain the faster saturation of the large-scale
field when $\Rm$ is increased.

\begin{figure}[t]
\centering
\includegraphics[width=\columnwidth]{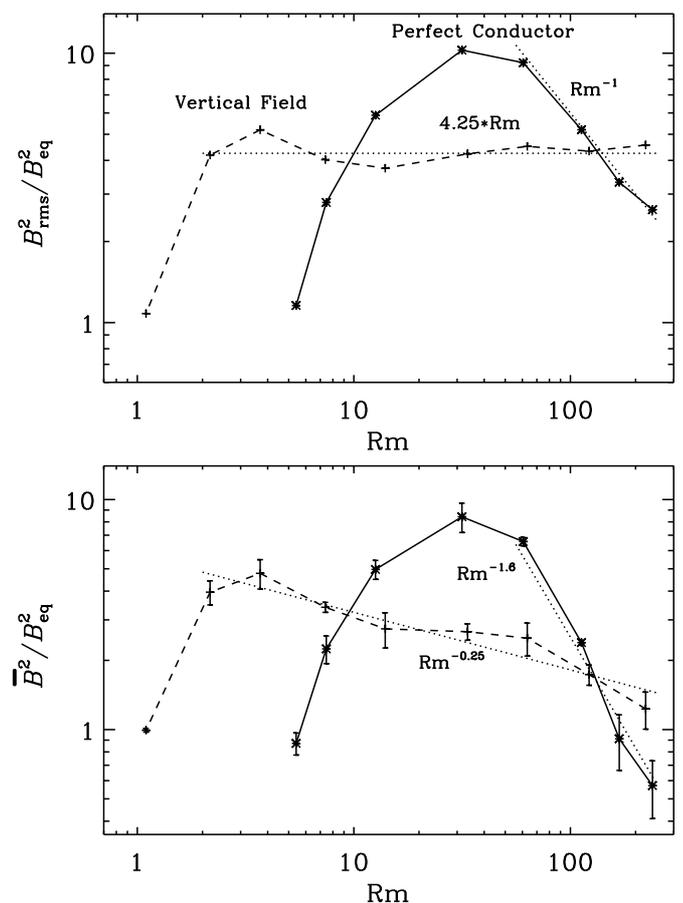}
\caption{Upper panel: rms-value of the total magnetic field in the
  saturated regime for runs with perfect-conductor (solid line) and
  vertical-field (dashed) boundaries. Lower panel: same as above but
  for the horizontally averaged mean magnetic field $\meanv{B}$. In
  both panels power laws fitting to the data at high $\Rm$ are shown
  for reference.}
\label{fig:pbrms}
\end{figure}

The saturation level of the total and mean magnetic fields for runs 
with open and
closed boundary conditions are shown in Fig.~\ref{fig:pbrms}.
We find that the total magnetic field in the runs with open boundary
conditions is roughly consistent with an $\Rm$-independent value. In
the perfect conductor case the total magnetic energy first increases 
up to $\Rm\approx30$ and then decreases
proportionally to $\Rm^{-1}$ for $\Rm\ga60$. Mean-field models taking
into account magnetic helicity evolution can also produce a maximum
for the saturation level at some intermediate $\Rm$ (e.g.\ Brandenburg
et al.\ \cite{BKMMT07}). We note that an $\Rm^{-1}$ dependence for the
mean field energy is indicative of catastrophic quenching. This would indeed
be the expected result for closed boundaries. However, in our case
only the total field energy shows the $\Rm^{-1}$ behaviour whereas 
the mean field
exhibits a much steeper (at least $\Rm^{-1.6}$) dependence. The
explanation for such a steep trend is as yet unclear. The data for the
mean magnetic field in the case of open boundaries also show a weak
decreasing trend consistent with a power law
$\Rm^{-0.25}$ as opposed to the expectation that the
saturation level is independent of $\Rm$.
However, the unexpected behaviour of the saturation level could simply 
be related to the fact that in mean-field models
true asymptotic behaviour may only commence at much larger values of $\Rm$
(Brandenburg et al.\ \cite{BCC09}).

\begin{figure}[t]
\centering
\includegraphics[width=\columnwidth]{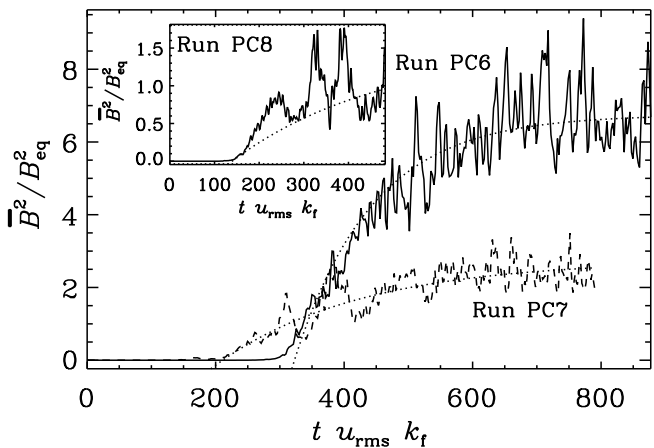}
\caption{Mean magnetic field as a function of time from Runs~PC6 (solid
  line), and PC7 (dashed). The inset shows the same for Run~PC8. 
  The dotted lines show a saturation
  predictor according to Brandenburg (\cite{B01}) with the
  microscopic values of $\eta$ and $k=2\pi/L_z$.}
\label{fig:pbbar}
\end{figure}

Our simulations with the highest magnetic Reynolds numbers 
and closed boundaries apparently do not
show a slow saturation behaviour; e.g.\ as in Fig.~\ref{fig:pbbar} where the
mean magnetic field and a saturation predictor (Brandenburg
\cite{B01}) proportional to $1-e^{-2\eta k^2(t-t_{\rm sat})}$ are shown
for Runs~PC6--PC8. 
Here we use $k=2\pi/L_z$ and $t_{\rm sat}$ is the time at
which the small-scale dynamo has saturated.
In the runs with intermediate $\Rm$ (PC6 and PC7) the saturation predictor 
is in fairly good agreement with the simulation results, whereas for 
Run~PC8 this is no longer the case.
This might be caused by additional contributions to $\meanv{B}$
whose effective value of $k$ is larger.
In fact, earlier simulations of forced turbulence with perfect conductor
boundary conditions (Brandenburg \& Dobler \cite{BD02}) showed that the
final configuration of the mean field can be established in steps,
but the time between different steps can still be resistively long.
This could explain why, in the highest-$\Rm$ simulations, the 
saturation level of the
mean magnetic field is lower than expected.
Whether this is also the case in the present simulations remains an 
open question.

\section{Conclusions}
We have studied the effects of magnetic boundary conditions on the excitation
and saturation of large-scale dynamos driven by turbulent convection,
shear, and rotation by means of numerical simulations. We find that
the critical magnetic Reynolds number is greater ($\Rm\approx5$) in
the case of closed boundaries (perfect-conductor) compared to
simulations with open (vertical-field) boundaries where the 
large-scale dynamo is
excited already for $\Rm\approx1$. A similar result is obtained from
mean-field models operating in the same parameter regime.
Furthermore, the relevant dynamo number is roughly three times greater
with closed boundaries. These effects are likely to explain the weak
large-scale dynamo action seen in the perfect-conductor runs in Paper~I.

The measured growth rate of the magnetic field is independent of the
microscopic resistivity when $\Rm$ is sufficiently above the
critical value. This is manifested by the approximately constant
growth rate in the intermediate $\Rm$ range in Fig.~\ref{fig:pgr}.
For $\Rm\ga30$, the small-scale dynamo is excited, and for
$\Rm\ga60$ it becomes dominant. The growth rate of the magnetic
field is then consistent with $\lambda\propto\Rm^{1/2}$ scaling, which is in
accordance with the results of Schekochihin et al.\ (\cite{Scheko04})
and Haugen et al.\ (\cite{{HBD04}}). The fact that the qualitative
behaviour of the growth rate of the magnetic field is similar for both
boundary conditions suggests that the origin of the large-scale dynamo is not
likely to be a process that is essentially nonlinear, as suggested by
Hughes \& Proctor (\cite{HP09}), but that it can be understood within
the framework of classical kinematic mean-field theory.

In the saturated state the energy of the total magnetic field remains
independent of $\Rm$ for
open boundaries and decreases as $\Rm^{-1}$ for closed 
boundaries. The
latter result is consistent with catastrophic quenching while the former
result suggests that magnetic helicity fluxes are efficiently driven
out of the system.
On the other hand, the energy of the mean field, taken here to be
represented by horizontal averages, decreases approximately as $\Rm^{-0.25}$ for
open and as $\Rm^{-1.6}$ for closed boundaries. It is not yet
clear why the mean fields tend to show a steeper decline than the total
field. It is possible that this declining trend levels off at a higher $\Rm$ and
that the magnetic Reynolds numbers in our simulations are still not 
large enough
(cf.\ Brandenburg et al.\ \cite{BCC09}). A similarly weak
$\Rm$-dependence has been observed in the cycle period of
$\alpha$-shear dynamos with isotropically forced turbulence
(K\"apyl\"a \& Brandenburg \cite{KB09}).

We find that, for intermediate values of $\Rm$, the large-scale 
magnetic field
saturates on a resistive time scale when closed boundaries are used.
However, with the largest $\Rm$ no clear signs of slow saturation are
observed. Earlier results using perfect-conductor boundaries have
shown that the mean field can evolve in steps (Brandenburg \& Dobler
\cite{BD02}; see also Brandenburg et al.\ \cite{BKMMT07}) which are
associated with a change of the large-scale magnetic field
configuration. We have not seen such a behaviour in our current
simulations but the existence of such events at a later stage cannot
be ruled out. Another possible explanation is that there are magnetic
helicity fluxes
occurring inside the domain which arise from the spatial gradients of
magnetic helicity (e.g.\ Covas et al.\ \cite{CTTB98}; 
Kleeorin et al.\ \cite{KMRS00}; Mitra et al.\
\cite{MCCTB10}).
However, a quantitative study of these effects requires more detailed
knowledge of the helicity fluxes and possibly an anisotropic formulation of the
magnetic $\alpha$ effect. These issues merit further investigation and
are beyond the scope of the present paper.

\begin{acknowledgements}
  The authors thank Anvar Shukurov for his detailed comments on the paper.
  The computations were performed on the facilities hosted by the CSC -- IT
  Center for Science in Espoo, Finland, administered by the
  Finnish ministry of education. We also wish to acknowledge the DECI
  -- DEISA network for granting computational resources to the
  CONVDYN project.
  Financial support from the Academy
  of Finland grants No.\ 121431 (PJK) and 112020 (MJK), as well as
  the Swedish Research Council grant 621-2007-4064
  and the European Research Council AstroDyn Research Project 227952 (AB)
  are acknowledged. The authors acknowledge the hospitality of Nordita
  during the program ``Solar and stellar dynamos and cycles''.
\end{acknowledgements}


\end{document}